\documentclass[twocolumn,showpacs,preprintnumbers,amsmath,amsfonts,prl]{revtex4}

\preprint{%to be submitted to {\em Physical Review Lett}
\hskip189pt LA-UR:03-0414}            % For preprint version

\usepackage{graphicx}% Include figure files
\usepackage{dcolumn}% Align table columns on decimal point
\usepackage{bm}% bold math

\usepackage{color}
\definecolor{red}{rgb}{1,0,0}
\definecolor{green}{rgb}{0,1,0}
\definecolor{blue}{rgb}{0,0,1}

\begin{document}

%\preprint{LA-UR:03-0414}

\title{The role of the lattice in the $\gamma \rightarrow \alpha$
phase transition of Ce: a high pressure neutron and x-ray
diffraction study}

\author{I.-K. Jeong}
\altaffiliation {} \email{jeong@lanl.gov}
\author{T. W. Darling}
\author{M. J. Graf}
\author{Th. Proffen}
\author{R. H. Heffner}
\affiliation{Los Alamos National Laboratory, Los Alamos, NM 87545,
USA.}
\author{Yongjae Lee}
\author{T. Vogt}
\affiliation{Physics Department, Brookhaven National Laboratory,
Upton, NY 11973-5000, USA}
\author{J. D. Jorgensen}
\affiliation{Materials Science Division, Argonne National
Laboratory, Argonne, IL 60439, USA}

%\date{\today}% It is always \today, today,
             %  but any date may be explicitly specified

\begin{abstract}
The temperature and pressure dependence of the thermal
displacements and lattice parameters were obtained across the
$\gamma \to \alpha$ phase transition of Ce using high-pressure,
high-resolution neutron and synchrotron x-ray powder diffraction.
The estimated vibrational entropy change per atom in the $\gamma
\to \alpha$ phase transition, $\Delta S^{\gamma - \alpha}_{\rm
vib} \approx (0.75 \pm 0.15)$k$_{\rm B}$, is about half of the
total entropy change. The bulk modulus follows a power-law
pressure dependence which is well described using the framework of
electron-phonon coupling. These results clearly demonstrate the
importance of lattice vibrations, in addition to the spin and
charge degrees of freedom, for a complete description of the
$\gamma \to \alpha$ phase transition in elemental Ce.
\end{abstract}

%\keywords{Suggested keywords}
\maketitle

Materials with electrons near the boundary between itinerant and
localized behavior continue to present a major theoretical
challenge to a complete description of their properties, including
multiple phases and anomalous thermodynamics. This is particularly
true in the 4$f$ and 5$f$ systems, where this boundary appears to
occur in or near the elements Ce and Pu, respectively
\cite{johansson;hf00}. In Pu, which possesses five allotropic
phases at ambient pressure, a partial localization of some of the
five 5$f$ electrons appears necessary to understand the higher
temperature phases~\cite{eriksson;jalc99}. Partial localization
may also be present in U compounds~\cite{Zwicknagl;jpcm03}.  Ce
metal is in principle simpler, possessing only a single 4$f$
electron, but still displays four different phases at ambient
pressure.  One of the most interesting and still not completely
understood phenomena in Ce is the isostructural (fcc) $\gamma
\rightarrow \alpha$ phase transition, which involves about 17\%
volume collapse at room temperature and pressure of roughly 0.8
GPa~\cite{koskenmaki;bk78}.

In the majority of theoretical models
~\cite{koskenmaki;bk78,johansson;pm74,johansson;prl95,
allen;prl82,allen;prb92,svane;prb99,jarlborg;prb97} the $\gamma
\to \alpha$ transition has been attributed to an instability of
the single 4$f^1$ electron. The earliest  models focused on charge
instability, while later models dealt with spin instability. The
promotional model postulates a transition from 4$f^1$5$d^1$6$s^2$
($\gamma$-phase) to 4$f^0$5$d^2$6$s^2$ ($\alpha$-phase), but is
inconsistent with the 4f binding energy and the cohesive energies
of other 5$d^2$6$s^2$ materials~\cite{johansson;pm74}. In the Mott
transition (MT) model~\cite{johansson;pm74,johansson;prl95} the
4$f$ electron in the $\gamma$ phase is localized and non-binding,
but  is itinerant and binding in the lower volume $\alpha$ phase.
The energy for the phase transition is provided by the kinetic
energy of the itinerant $f$ electron. In the Kondo-volume-collapse
(KVC) model~\cite{allen;prl82,allen;prb92} the 4$f$ electron is
assumed to be localized in both the $\gamma$ and $\alpha$ phases,
and the phase transition is driven by the Kondo spin fluctuation
energy and entropy within the context of the single-impurity
Anderson model. These early models ignored altogether an explicit
treatment of the lattice degrees of freedom; even the lattice
entropy is not considered.  More recent treatments
\cite{jarlborg;prb97,jarlborg;rpp97,wang;prb00} include both the
lattice and spin entropies, but still do not deal explicitly with
the consequences of electron-lattice coupling despite the large
volume collapse at the transition.

In rare-earth compounds the electron-phonon coupling can be
important because the ionic radii of different valences often
differ by over 10\%~\cite{varma;rmp76}. For example, the anomalous
phonon properties in mixed-valent Sm$_{0.75}$Y$_{0.25}$S and
YbInCu$_4$ have been ascribed to a strong coupling to its valence
instability~\cite{mook;prb78,entel;prl79,kindler;prb94,sarrao;prb98}.
In addition, it has been shown that lattice vibrational
contributions renormalize the two essential parameters of the
Anderson model: the hybridization energy  and energy of 4$f$
state~\cite{entel;zpb79,entel;prl79}. In $\gamma$-Ce, a
comparison~\cite{stassis;prb79} of phonon dispersion curves with
those of thorium shows that the longitudinal branches of Ce are
much softer than one would expect from the Lindemann
rule~\cite{linde;pz10}, which accounts for the differences in
interatomic distance, mass, and melting temperature of these
elements. This relative softening, accompanied by a change in 4$f$
localization (see conclusions), suggests an important role for
electron-phonon coupling, as in Sm$_{0.75}$Y$_{0.25}$S, where the
[$\xi\xi\xi$] longitudinal phonon branch is also soft compared to
transverse branches~\cite{mook;prb78}.

In this Letter, we further illuminate the role of the lattice in
the $\gamma \to \alpha$ transition and present the first
(remarkably) neutron diffraction experiment under pressure in
elemental Ce. We measured thermal displacements of Ce in the
$\gamma$ and $\alpha$ phases as a function of temperature at
constant pressures ($P\sim$ 0.4 and 0.5 GPa), and as a function of
pressure at 300 K. The vibrational entropy change between the
$\gamma$ and $\alpha$ phases, obtained from the thermal
displacements using the Debye approximation,  accounts for about
half of the total entropy change and is thus non-trivial.
Furthermore, the pressure dependences of the bulk modulus and
thermal displacements are consistent with the explicit inclusion
of electron-phonon coupling~\cite{bergman;prb76,bruce;bk81}. These
results suggest that lattice dynamics and electron-phonon
coupling, in addition to electronic and spin instabilities, are
essential ingredients for a complete understanding of the $\gamma
\to \alpha$ phase transition in elemental Ce .

A high-purity (99.99 \%) polycrystalline, cylindrical ingot of Ce
was obtained from the Ames Laboratory. In order to stabilize the
$\gamma$ phase at room temperature, the Ce rod was sealed in a
fused silica tube with low pressure ($<$ 1 atm) Argon gas and
placed in a furnace for 48 hours at 423~K, and then furnace cooled
over 6 hours.  Powder diffraction patterns of the $\gamma$ and
$\alpha$ phases were collected at room temperature in the pressure
range up to 0.85~GPa on NPD at the Los Alamos Neutron Science
Center (LANSCE) and up to 3.5 GPa on the X-7A beamline at the
National Synchrotron Light Source (NSLS), Brookhaven National
Laboratory~\cite{vogt;prb01}. Temperature dependent measurements
at constant pressure ($P\sim$ 0.4 and 0.5 GPa) were performed on
SEPD at the Intense Pulsed Neutron Source (IPNS), Argonne National
Laboratory. For the neutron powder diffraction measurements,  high
pressure Al cells were used for the pressure range up to 0.85~GPa.
Contamination due to scattering from the cell was minimized using
a Gd-based internal shielding, with spaces for the incoming,
outgoing and $\pm 90^\circ$ scattered beam. Details of the cell
designs are described elsewhere~\cite{jorgensen;pc90,darling;02}.

\begin{figure}[!tb]\vspace{-0cm}
\includegraphics[angle=0,scale=1.2]{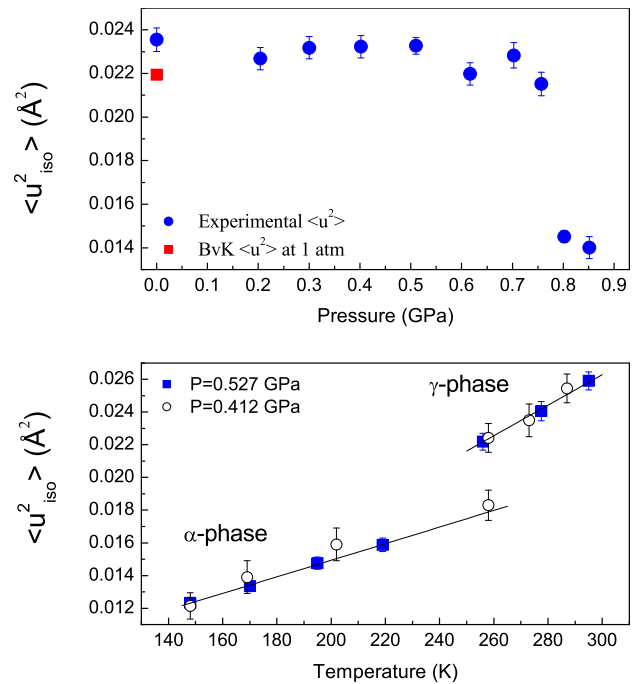}
\caption{(Upper panel)
 Isotropic thermal displacements of Ce vs
pressure at 300 K. Symbols: circles are experimental data; the
square is calculated from Born-von K\'{a}rm\'{a}n forces at
ambient conditions~\cite{stassis;prb79,jeong;prb03}. Across the
phase transition the thermal displacement drops about 30\%-40\%.
(Lower panel)
 Thermal displacements vs temperature at 0.527 GPa
(closed squares) and 0.412 GPa (open circles). Data at 0.527 GPa
were collected with decreasing temperature and data at 0.412 GPa
were collected with increasing temperature. Data at T=260 K has a
mixed $\gamma$ and $\alpha$ phases. Solid lines are fits to Eq.
(\ref{eq;debye}) with Debye temperatures
$\Theta^{\gamma}_D$=104(3) K and $\Theta^{\alpha}_D$=133(3) K in
$\gamma$ and $\alpha$ phases, respectively.}
\label{fig;ce_thermal_displacement}
\end{figure}

Fig.~\ref{fig;ce_thermal_displacement} shows the isotropic thermal
displacements ~$\langle u_{iso}^2 \rangle$ of Ce as a function of
pressure (upper panel) and temperature (lower panel) obtained from
the Rietveld refinement using GSAS~\cite{gsas} and
EXPGUI~\cite{expgui}.
The notable features in the upper panel are the very weak pressure
dependence of $\langle u_{iso}^2 \rangle$ below 0.7 GPa and the
pronounced drop of $\langle u_{iso}^2\rangle$ ($\approx 30-40\%$)
at the phase transition, which indicates a significant stiffening
of the lattice.   The square symbol is the thermal displacement
calculated from measured phonon dispersion
curves~\cite{stassis;prb79} using the Born-von K\'{a}rm\'{a}n
force model at ambient conditions~\cite{jeong;prb03}; the
agreement with our measurement is quite good. One finds that these
measurements yield essentially the same relative change of thermal
displacement (u$^2_{\gamma}$/u$^2_{\alpha}$) at the phase
transition as the constant pressure measurements described next.
In Fig.~\ref{fig;ce_thermal_displacement} (lower panel) the closed
squares represent  measurements performed at 0.527 GPa with
decreasing temperature and open circles at 0.412 GPa with
increasing temperature, respectively. At T=260 K (open circle) we
had mixed $\gamma$ and $\alpha$ phases and were able to obtain the
thermal displacements of both phases.
The solid lines are fits using
Eq.(\ref{eq;debye})~\cite{lawson;jlcm91} with Debye temperatures
$\Theta^{\gamma}_D$=104(3) K and $\Theta^{\alpha}_D$=133(3) K,
\begin{equation}
\langle u^2 \rangle_{\rm measured} = \langle u^2 \rangle_{\rm
offset}+\frac{3 \hbar}{M \omega_D}
\biggl[\frac{1}{4}+\biggl(\frac{T}{\Theta_D}\biggr)^2 \Phi_1
\biggr],
 \label{eq;debye}
\end{equation}
where $\Phi_1 = \int_{0}^{\Theta_D/T} x (e^x-1)^{-1}~dx$,
$\Theta_D$ (=$\hbar \omega_D/k_B$) is the Debye temperature, and
$\langle u^2 \rangle_{\rm offset}$ is a constant offset.

At high temperatures, $T > \Theta_D$, the vibrational entropy can
be approximated by $S_{\rm vib} \approx 3Nk_B[1+{\rm
ln}(T/\Theta_0) + \dots ]$, where $\Theta_0$ is the logarithmic
phonon moment~\cite{wallace;bk02}. Approximating $\Theta_0$ by
$\Theta_D$ ($\Theta_D \approx \Theta_0 e^{1/3}$) yields the
vibrational entropy change per atom, $\Delta S^{\gamma -
\alpha}_{\rm vib} \equiv S^{\gamma}_{\rm vib} - S^{\alpha}_{\rm
vib}$, which can  be expressed as~\cite{grimvall;bk99}
\begin{equation}
 \Delta S^{\gamma - \alpha}_{\rm vib} \approx
 3 k_B\ln
  {\Theta^{\alpha}_D / \Theta^{\gamma}_D}
 .
\label{eq;entropy-vib}
\end{equation}
Using Eq.~\ref{eq;entropy-vib} we obtain $\Delta S^{\gamma -
\alpha}_{\rm vib} \approx 3 k_{\rm B} \ln\Big( \frac{133\pm
3~K}{104\pm 3~K} \Big) \approx (0.75 \pm 0.15) k_{\rm B}$.
This change is roughly half of the total entropy change, $\Delta
S^{\gamma - \alpha}_{\rm tot} \approx 1.54  k_{\rm B}$
\cite{koskenmaki;bk78}, which follows from the latent heat or the
Clausius-Clapeyron relation, $dP/dT = \Delta S^{\gamma -
\alpha}_{\rm tot} / \Delta V^{\gamma - \alpha}$.
This large change in vibrational entropy is qualitatively
consistent with sound speed measurements
\cite{voronov;spd60,voronov;spjetp79} in pure Ce
($\Theta_{\gamma}\approx 137\ {\rm K}$ and $\Theta_{\alpha}\approx
154\ {\rm K}$),
but not with the measurements of the phonon density of states of
the Ce$_{0.9}$Th$_{0.1}$ alloy at 150~K ($\gamma$-phase) and at
140~K ($\alpha$-phase) by Manley {\it et al.}~\cite{manley;prb03},
which showed little difference between $\gamma$ and $\alpha$
phases. We suspect that thorium atoms in Ce$_{0.9}$Th$_{0.1}$
alloys  modify the elastic properties of Ce, as is observed in
Mg-doped Ce alloys, where a few percent Mg significantly affect
the Young's modulus, particularly in the $\alpha$
phase~\cite{clinard;jap69}.

\begin{figure}[!tb]
\includegraphics[angle=0,scale=1.1]{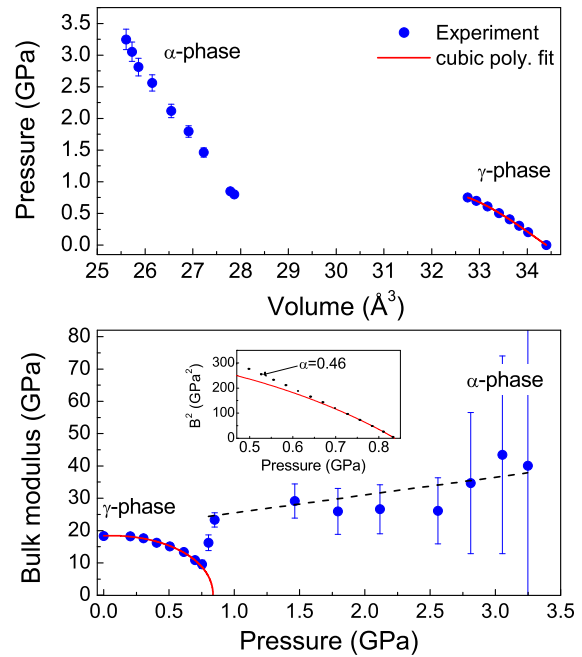}
\caption{(Upper panel)
 $P$-$V$ room temperature isotherm.
The solid line is a cubic polynomial fit to the $\gamma$ phase of
Ce. (Lower panel) Bulk modulus vs pressure. The symbols are
calculated from finite differences. The solid line is obtained
from the polynomial fit to the $\gamma$ phase $P$-$V$ data in
upper panel. The dashed line in the $\alpha$ phase is from Olsen
{\it et al.}~\cite{olsen;physicab85}. Inset: The dashed line is a
fit to the solid line near the transition (0.7 GPa$\leq$ P $\leq$
0.837 GPa), where $B_T^2 \sim |P-P_{c}|^{2\alpha}$ with $P_{c}
\approx 0.83~{\rm GPa}$ and $\alpha \approx 0.46$.}
\label{fig;ce_hp_bk_latt}
\end{figure}

We now focus on the softening of the isothermal bulk modulus $B_T$
with increasing pressure in the $\gamma$ phase. The upper panel of
Fig.~\ref{fig;ce_hp_bk_latt} presents the measured $P$-$V$ data.
The closed circles in the lower panel of
Fig.~\ref{fig;ce_hp_bk_latt} were obtained by finite differences
of the $P$-$V$ data, $B=-V dP/dV \approx -V \Delta P / \Delta V$.
Here the errors are estimated using the errors in applied
pressures and lattice parameters. Lacking an equation of state
that spans the $\gamma \to \alpha$ transition, we fit the $\gamma$
phase $P(V)$ data to a cubic polynomial for a more accurate
description of $B_T$, shown as solid line in the upper panel of
Fig.~\ref{fig;ce_hp_bk_latt}. The $B_T(P)$ derived from this fit
is shown as solid line in the lower panel of
Fig.~\ref{fig;ce_hp_bk_latt}. The dashed line in the $\alpha$
phase is from Olsen {\it et al.}~\cite{olsen;physicab85}. We note
that our results are in good agreement with the adiabatic bulk
modulus $B_S$ obtained from ultrasound measurements by Voronov
{\it et al.}~\cite{voronov;spd60, voronov;spjetp79}. (The ratio
between the isothermal and isentropic bulk modulus is
B$_T$/B$_S$=C$_P$/C$_V \simeq 1$ for Ce~\cite{gschneidner;bk64} at
room temperature.) Fig.~\ref{fig;ce_hp_bk_latt} shows that $B_T$
decreases with increasing pressure below $P < P_{c}$, and that its
smooth extrapolation in the $\gamma$ phase vanishes at $P_{c}$,
indicating an elastic instability of the lattice. The
discontinuity of $B_T$ in the region $P \approx P_{c}$ reflects
the phase instability and the first-order nature of the
transition.

The softening of the bulk modulus with increasing pressure in the
$\gamma$ phase is directly related to the softening of the
$C_{11}$ elastic constant. In a cubic lattice the bulk modulus is
$B = [3C_{11}-4C^*+P]/3$, where $P$ is the applied hydrostatic
pressure and $C^*=(C_{11}-C_{12})/2$ is a shear
modulus~\cite{wallace;bk98}. $C^*$ is known to be pressure
insensitive in the $\gamma$
phase~\cite{voronov;spd60,voronov;spjetp79,bergman;prb76}.
Therefore, the softening of the bulk modulus in the $\gamma$ phase
is a direct consequence of the softening of $C_{11}$. This result
is consistent with the softening of the measured longitudinal
sound speed with increasing pressure in
$\gamma$-Ce~\cite{voronov;spd60,voronov;spjetp79}, and is related
to the softening of longitudinal phonons~\cite{stassis;prb79} at
ambient pressure mentioned above.

The pressure dependence of the thermal displacement and $B_T$ of
Ce can be described within the model of order-parameter-strain
field coupling, e.g., $\lambda e Q^2$, where $e$ is the strain,
$Q$ is a scalar order parameter, and $\lambda$ is the coupling
constant. Bergman and Halperin showed that in this model a
continuous phase transition is preempted by an elastic
instability, causing a renormalization of the bulk
modulus~\cite{bergman;prb76}. (In rare-earth compounds,  the order
parameter $Q$ has been defined as a relative change of the lattice
constant ($\Delta a/a$)~\cite{entel;zpb79,entel;prl79}, which is
proportional to the 4$f$ occupation number
$n_f$~\cite{franceschi;prl69, cornelius;prb97}. Because the strain
couples to the square of the order parameter, it constitutes a
``secondary'' order parameter~\cite{bruce;bk81}.) In this
framework the thermal displacement behaves as $\langle u^2\rangle
\sim E_1+E_2 p+E_3 p^{1-\alpha}$, where $p\equiv \mid P-P_{c}
\mid$, E$_i$ are constants and $\alpha$ is the specific heat
critical exponent~\cite{meissner;prb75}. The bulk modulus vanishes
as $B\sim p^{\alpha}$. In mean-field theory the exponent is
$\alpha=0$, while for Gaussian fluctuations one expects
$\alpha=\frac{1}{2}$ \cite{bruce;bk81}.

As shown in Figs.~\ref{fig;ce_thermal_displacement} and
\ref{fig;ce_hp_bk_latt} our experimental data exhibit  the
features of such a model. Our pressure-dependent measurements at
300 K show no divergence of the thermal displacement, instead,
they exhibit a sharp drop across the first-order transition. In
addition, we find a continuous decrease of the bulk modulus
towards zero as the pressure approaches $P_{c}$ with a simple
scaling behavior, $B_T \sim p^\alpha$ (see inset in lower panel of
Fig.~\ref{fig;ce_hp_bk_latt}). Near the transition we extract the
exponent $\alpha \approx 0.46$ and the critical pressure $P_{c}
\approx 0.83$ GPa. Of course, $\alpha$ is not a truly measured
``critical exponent'' because (1) we have only a few measured data
points below the transition, and (2) the descent of $B_T$ towards
zero is clearly preempted by the first-order phase transition, as
seen by the large volume collapse.

In conclusion, these new experiments and analysis clearly
demonstrate that vibrational entropy plays a significant role in
stabilizing $\gamma$-Ce, accounting for about half of the total
change in entropy at the $\gamma \to \alpha$ transition. The
pressure dependence of the thermal displacement and bulk modulus
of Ce also strongly suggest that  electron-phonon coupling plays
an important role in  the $\gamma \to \alpha$ transition.
Finally, we note that our results mean that a complete
understanding of Ce and its fascinating $\gamma \to \alpha$
transition must take into account the important interactions
between  the spin, charge \emph{and lattice} degrees of freedom,
and that this relates closely to the competition between
localization
and itinerancy in $f$-electron materials.\\

\acknowledgments

We gratefully acknowledge discussions with A. C. Lawson, Y. Bang,
J. M. Wills, D. Hatch and are thankful to A.C.L and S. Short for
helping with the data collection. Work at LANL was carried out
under the auspices of the US Department of Energy. Part of the
data was collected at the Manuel Lujan, Jr.\ Neutron Scattering
Center and the IPNS at Argonne, which are  national user
facilities funded by the U.S. Department of Energy.
%Office of Basic Energy Sciences.
Research was carried out in part at the National Synchrotron Light
Source, Brookhaven National Laboratory, which is supported by the
U.S. Department of Energy, Division of Materials Sciences and
Division of Chemical Sciences, under Contract No.
DE-AC02-98CH10886.

%\bibliography{C:/revtex4/bib/1995,%
%              C:/revtex4/bib/1996,%
%              C:/revtex4/bib/1997,%
%              C:/revtex4/bib/1998,%
%              C:/revtex4/bib/1999,%
%              C:/revtex4/bib/2000,%
%              C:/revtex4/bib/2001,%
%              C:/revtex4/bib/2002,%
%              C:/revtex4/bib/2003,%
%}

%
%\bibliography{apssamp}
%\bibliographystyle{/u24/billinge/tex/bib/aip}

\end{document}